\documentclass[letterpaper,fleqn,usenatbib]{mnras}

\usepackage{enumerate}

\def\eg{{\it e.g.}}
\def\etal{{\it et al.}}

\def\ie{{\it i.e.}}

\def\Msun{M$_\odot$}

\def\pmb#1{\setbox0=\hbox{$#1$}%
  \kern-0.25em\copy0\kern-\wd0
  \kern.05em\copy0\kern-\wd0
  \kern-0.025em\raise.0433em\box0}
\def\spmb#1{\setbox1=\hbox{${\scriptstyle #1}$}%
  \kern-0.25em\copy1\kern-\wd1
  \kern.05em\copy1\kern-\wd1
  \kern-0.025em\raise.0433em\box1}

\usepackage{graphicx}	

\title[Warps in disc galaxies]{Internally-driven warps in disc galaxies}

\author[Sellwood \& Debattista]
{J. A. Sellwood,$^{1}$\thanks{E-mail:sellwood@as.arizona.edu}
and
{Victor P. Debattista,$^{2}$\thanks{E-mail:vpdebattista@gmail.com}}
\\
$^1$Steward Observatory, University of Arizona, 933 N Cherry Ave, Tucson AZ 85722, USA \\
$^2$Jeremiah Horrocks Institute, University of Central Lancashire, Preston, PR1 2HE, UK}

\pubyear{2021}

\begin{document}
\label{firstpage}
\pagerange{\pageref{firstpage}--\pageref{lastpage}}
\maketitle

\begin{abstract}
Any perturbation to a disc galaxy that creates a misalignment between
the planes of the inner and outer disc, will excite a slowly evolving
bending wave in the outer disc.  The torque from the stiff inner disc
drives a retrograde, leading-spiral bending wave that grows in
amplitude as it propagates outward over a period of several Gyr.  This
behaviour creates warps that obey the rules established from
observations, and operates no matter what the original cause of the
misalignment between the inner and outer disc.  The part of the disc
left behind by the outwardly propagating wave is brought into
alignment with the inner disc.  Here we confirm that mild warps in
simulations of disc galaxies can be excited by shot noise in the halo,
as was recently reported.  We show that the quadrupole component of
the noise creates disc distortions most effectively.  Bending waves
caused by shot noise in carefully constructed equilibrium simulations
of isolated galaxies are far too mild to be observable, but
perturbations from halo substructure and galaxy assembly must excite
larger amplitude bending waves in real galaxies.

\end{abstract}

\begin{keywords}
galaxies: spiral ---
galaxies: evolution ---
galaxies: structure ---
galaxies: kinematics and dynamics ---
\end{keywords}


\section{Introduction}
\label{sec.intro}
Warps in the neutral hydrogen layer of the Milky Way
\citep[\eg][]{Oort58, Levine06} and other disc galaxies
\citep[\eg][]{Sancisi76, Bosma91} are now established as a common
feature of galaxies.  The visible disc can also be twisted to a
smaller extent; where both are observed, the distortion of the stellar
layer may follow that of the gas in some galaxies, \eg\ UGC~7170
\citep{Cox96}, which is strong evidence that its warp is a
gravitationally-driven phenomenon.  \citet{SS06} successfully modeled
the ``prodigious'' warp of NGC~4013 as a gravitationally-driven
bending wave.  However, there are also galaxies, \eg\ NGC~4565
\citep{RS+14} and perhaps also the Milky Way \citep{Reyle09, Chen19},
that have a second type of warp in which the gas, and even some young
stars, are outside the stellar layer defined by the old disc. The more
likely explanation for the misaligned gas and young stars in these
cases is off-axis gas accretion.  Our focus in this paper is on warps
of the first kind.

\citet{Briggs90} identified three rules that were followed by all the
well observed galaxies in his sample:
\begin{enumerate}
\item The H\,I layer typically is coplanar inside radius $R_{25}$, the
  radius where the B-band surface brightness is $25\,\hbox{mag
    arcsec}^{-2}$, and the warp develops between $R_{25}$ and
  $R_{26.5}$ (the Holmberg radius).
\item The line of nodes (LoN) is roughly straight inside $R_{26.5}$.
\item The LoN takes the form of a loosely-wound leading spiral outside
  $R_{26.5}$.
\end{enumerate}

\citet{KW59} first drew attention to the theoretical challenge
presented by disc warps, emphasizing that the gravitational stresses
in a twisted disk would be too weak to resist the winding of the warp
by differential precession.  As the theory of gravitationally-driven
warps was thoroughly reviewed by \citet{Sell13}, we here briefly
summarize two aspects that are of relevance to the present paper.

\subsection{Bending waves in discs}
\label{sec.bends}
The frequency of bending waves of wavenumber, $k = 2\pi/\lambda$, in a
uniform, razor-thin, sheet of stars having a mass surface density
$\Sigma$ is given by the dispersion relation
\begin{equation}
\omega^2 = 2\pi G \Sigma|k| - \sigma_x^2k^2 + \nu_{\rm ext}^2,
\label{eq.disrel}
\end{equation}
where $\sigma_x$ is the stellar velocity dispersion in the direction
of the wave, \ie\ perpendicular to the wave crests, and $\nu_{\rm
  ext}^2 = |\partial^2\Phi_{\rm ext}/\partial z^2|_{z=0}$ \citep{BT08}
is the squared vertical oscillation frequency due to mass, such as the
bulge and halo, that is not in the sheet.  Bending waves are stable
when $\omega^2 >0$, which requires $\lambda > \sigma_x^2/G\Sigma$ if
$\nu_{\rm ext} = 0$.

The WKB approximation assumes equation (\ref{eq.disrel}) holds for
tightly wrapped bending waves in a thin disc, which requires
wavelengths short enough that their curvature can be neglected, or
ideally that $|kR|\gg1$.  As for spiral waves, the tight winding
approximation allows one to write the forcing frequency $\omega =
m(\Omega_p - \Omega_c)$ for an $m$-armed bending wave.  Here
$\Omega_p$ is the angular pattern speed of the wave and $\Omega_c$ is
the local circular frequency.  Even though bending waves in real
galaxies are not short-wavelength tightly-wrapped disturbances, this
analysis is believed to give some qualitative indication of their
behaviour.  With these approximations, the group velocity of a packet
of bending waves is \citep{To83}
\begin{equation}
v_g \equiv {\partial \omega \over \partial k} = {\hbox{sgn}(k)\pi
  G\Sigma - k\sigma_R^2 \over m(\Omega_p - \Omega_c)}.
\label{eq.vgroup}
\end{equation}
The convention adopted in this analysis is that $k<0$ for leading
waves, and therefore the group velocity of slow ($\Omega_p\ll
\Omega_c$) leading waves is positive, or radially outward in the disc.
Note that an outwardly decreasing surface density of the disc has two
consequences: first, $v_g$ slows as a wave packet moves outward and
second, by analogy with whips, wave action conservation requires its
amplitude also to rise.

\subsection{Warps}
\label{sec.warps}
\citet{HT69} studied the bending modes of isolated, razor-thin disks
with no random motion.  They found that the Maclaurin disc, which has
a sharp outer edge, had a discrete spectrum of bending modes, but only
two trivial modes existed in discs that had fuzzy edges: a uniform
vertical displacement and a simple tilt of the entire disc.  The
importance of a fuzzy edge is that bending waves travel ever more
slowly as they approach the edge and could never reflect to create the
standing wave required for a mode.  Although this conclusion holds
only for discs lacking random motion \citep[see][for a
  counter-example]{Sell96}, the effect of galaxy halos has gradually
emerged as a greater obstacle to the idea that warps could be
long-lived modes of discs.

\citet{SC88} developed an idea originally due to \citet{DS83} that
galactic warps arise because the disc of a galaxy is misaligned with
the principal plane of a rigid oblate (or prolate) dark matter halo.
The twist of the disk, which they described as a modified tilt mode,
arose where the dominant contribution to the gravitational field
transitioned from the disc to the misaligned halo.  Unfortunately,
their assumption that the halo would behave as a rigid mass
distribution does not hold; \citet{KD95}, \citet{BJD98} and
\citet{SS06} all reported that the orbits of live halo particles
quickly adjust to the total potential of the disc plus halo to
eliminate the hypothesized misalignment.

The most promising idea for warp creation originated with \citet{OB89}
who proposed that the angular momentum of late infalling material as a
disc galaxy is assembled would probably be misaligned with the angular
momentum vector of the original disc.  This idea was explored by
\citet{QB92}, and the gravitational attraction of a misaligned annulus
of matter was modelled by \citet{JB99} and \citet{SS06}.  These last
authors, in particular, reported anti-symmetric ($m=1$) bending waves
in the disc that would last not forever, but for long enough that it
is likely that any galaxy would suffer further accretion of material
having a different angular momentum vector before the first warp had
decayed.

Alternatively the gas warp is the infalling material itself.  In
massive galaxies, gas is believed to reach the disc via hot accretion,
in which the intergalactic gas is shock-heated as it crosses the
halo's virial radius \citep[\eg][]{FE80, Brook04, Keres05,
  Robertson06}. Cosmological simulations show that the angular
momentum of the resulting hot gas corona is randomly oriented with
respect to that of the disc \citep{vdBosch02, Velliscig15, Stevens17,
  Obreja19}.  Consequently, cooling coronal gas reaches the disc
misaligned, forming a warp \citep{Roskar10}, which also drives a slow
tilting of the disc \citep{Earp19}. This picture naturally explains
why gaseous warps are traced only by younger stars in the Milky Way
\citep{Chen19} and NGC~4565 \citep{RS+14}.

\subsection{Present study}
With this background, \citet[][hereafter CW17]{CW17} were surprised to
find long-lived coherent bending waves in their two stellar dynamical
simulations of isolated Milky Way models.  Here we create and evolve
similar simulations which show that CW17 correctly attributed the
excitation of disc distortions to particle noise in the halo.  We also
show that these disturbances created coherent gravitationally-driven
bending waves in the disc that are easily understood.

The mechanism for the slowly evolving warps that we, and earlier
\citet{SS06}, identify is quite general and requires only that the
inner disc be misaligned with the outer disc, no matter what created
the misalignment.

\section{Technique}
\label{sec.methods}

\subsection{Disc-bulge-halo model}
\label{sec.DBHmodel}
We adopt the model for a disc galaxy embedded in a bulge and halo
that was previously used by \citet{Sell21}.  The dense bulge inhibits bar
instabilities \citep{To81}, and the disc supports mild spirals.

\begin{figure}
\includegraphics[width=.9\hsize,angle=0]{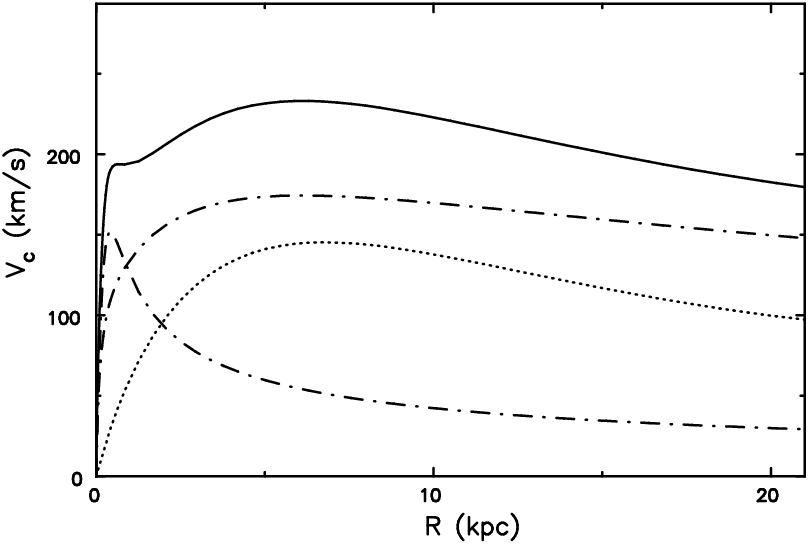}
\caption{The initial rotation curve of our galaxy model (solid curve)
  measured from the particles.  The disc contributes the dotted line,
  while the dot-dash lines indicate the contributions of the bulge and
  compressed halo.}
\label{fig.DBH-rc}
\end{figure}

\begin{itemize}
\item The exponential disc has the surface density
\begin{equation}
\Sigma(R) = {M_d \over 2\pi R_d^2} \exp(-R/R_d),
\label{eq.surfd}
\end{equation}
where $M_d$ is the mass of the notional infinite disc and $R_d$ is the
disc scale length.  The vertical density profile of the disc is
Gaussian with a scale $0.1R_d$ and, in this work, we truncate the disc
at $7R_d$.  Since we are interested in warps in the outer disc,
employing equal mass particles to represent such an extensive disc
would result in merely $\sim1$\% of the disc particles in the range
$6<R/R_d<7$.  In order to employ a larger fraction of the particles in
the outer disc, we set the masses of the particles $\propto
R\exp(-R/R_d)/R_d$, based on their initial radii, to achieve equal
numbers in equal radial bins while preserving the exponential density
profile (\ref{eq.surfd}).

\item We use a dense Plummer sphere for the bulge, which has a mass of
  $0.1M_d$ and core radius $a=0.1R_d$, and adopt the isotropic
  distribution function (DF) given by \citet{Dejo87}.  The bulge is
  dense enough that it dominates the central attraction in the inner
  part of our model, and the analytic DF is close to equilibrium
  despite the presence of the disc and halo.

\item We start from a \citet{Hern90} model for the halo that has the
  density profile
\begin{equation}
\rho(r) = {M_h b \over 2\pi r ( b + r )^3},
\label{eq.halod}
\end{equation}
where $M_h$ is the total mass integrated to infinity, and $b$ is a
length scale.  We choose $M_h = 5M_d$ and $b=4R_d$.  Naturally, the
isotropic DF that Hernquist derived for this isolated mass
distribution would not be in equilibrium when the disc and bulge are
added, so we use the adiabatic compression procedure described by
\citet{SM05} that starts from the DF given by Hernquist and uses the
invariance of both the radial and azimuthal actions to compute a
revised density profile and DF as extra mass is inserted.  The revised
DF has a slight radial bias but remains spherical.  We also
apply an outer cutoff to the selected particles that excludes any with
enough energy ever to reach $r>10b$, which causes the halo density to
taper smoothly to zero at that radius.
\end{itemize}
The resulting rotation curve of the model is illustrated in
Figure~\ref{fig.DBH-rc}, where we have scaled the model, both here and
throughout the paper, so that $R_d = 3\;$kpc and the unit of time
$(R_d^3/GM_d)^{1/2} = 12\;$Myr, which implies a disc mass $M_d =4.17
\times 10^{10}\;$\Msun.

We adopt a constant value of $Q=1.5$ at all radii to determine the
radial velocity dispersion of the disc particles.  Although the disc
is quite massive, the high value of the epicyclic frequency, $\kappa$,
near the centre in particular implies random velocities are modest
everywhere, and the Jeans equations in the epicyclic approximation
\citep{BT08} yield an excellent disc equilibrium.

\subsection{Numerical method}
The particles in our simulations move in a 3D volume that is spanned
by two separate grids; a cylindrical polar mesh and a much larger
spherical grid.  The self-gravitational attractions are calculated at
grid points and interpolated to the position of each particle.  The
disc particles are initially assigned to the polar grid, while the
spherical grid is used for the bulge and halo particles.  Naturally,
all particles are attracted by all others at every step.  A full
description of our numerical procedures is given in the on-line manual
\citep{Sell14} and the code itself is available for download.

Table \ref{tab.DBHpars} gives the default values of the numerical
parameters for the simulations presented in this paper, and cases
where they are changed are noted in the text.  It is easy to restrict
the spherical harmonics that contribute to the field of the halo
particles.  Also, the sectoral harmonics that contribute to the field of
the disc particles on a polar grid, since that part of the
gravitational field is a convolution of the mass density with a Green
function that is most efficiently computed by Fourier
transforms.

\begin{table}
\caption{Default numerical parameters} 
\label{tab.DBHpars}
\begin{tabular}{@{}ll}
Polar grid size & 175 $\times$ 256 $\times$ 125 \\
Grid scaling & $R_d= 10$ grid units \\
Vertical spacing & $\delta z = 0.02R_d$ \\
Active sectoral harmonics & $0 \leq m \leq 4$ \\
Softening length & $R_d/10$ \\
Spherical grid & 501 shells \\
Active spherical harmonics & $0 \leq l \leq 4$ \\
Number of disc particles & $10^7$ \\
Number of halo particles & $10^7$ \\
Number of bulge particles & $10^6$ \\
Basic time-step & $(R_d^3/GM)^{1/2}/320$ \\
Time step zones & 6 \\
\end{tabular}
\end{table}

\subsection{Other details}
We define a set of radii $\{R_k\}$ that are spaced every $0.2R_d$ to
separate the disc into radial bins and measure bending distortions in
each bin by forming the transform
\begin{equation}
Z_m(R_{k+n},t) = {\sum_j w_n\mu_j z_je^{im\phi_j} \over \sum_j w_n \mu_j}, \qquad n=0,1
\label{eq.zanl}
\end{equation}
where $\mu_j$ is the mass and $(R_j,\phi_j,z_j)$ are the cylindrical
polar coordinates of the $j$-th particle at time $t$.  The weights
$w_0$ and $w_1$ share the particle's contribution linearly according
to its radius $R_j$ between the rings at $R_k$ and $R_{k+1}$.

In order to measure frequencies of $m$-fold symmetric bending waves
that we assume have coherent frequencies across a broad swath of the
disc, we fit these data using the procedure described by \citet{SA86}.
Here, we assume constant amplitude waves that are the real part of
\begin{equation}
{\cal Z}_m(R_k,\phi,t) = {\cal A}_m(R_k)e^{i(m\phi - \omega t)},
\label{eq.mode}
\end{equation}
where the frequency $\omega = m\Omega_p$ and $\Omega_p$ is the pattern
speed.  We will be concerned exclusively with $m=1$ warps, and
generally we fit for two co-existing coherent waves.  The complex
function for each disturbance ${\cal A}_m(R_k)$, which is independent
of time, describes the radial variation of the vertical displacement
and phase of the wave, which would be a normal mode of the system if
${\cal A}_m(R_k)$ is strictly independent of time for all $\{R_k\}$,
or it may simply be a bending wave propagating slowly across the disc.
It is possible to fit for growing displacements by allowing $\omega$
to be complex, but we do not do that here.

\section{Results}
\label{sec.res}
\begin{figure}
\includegraphics[width=.9\hsize,angle=0]{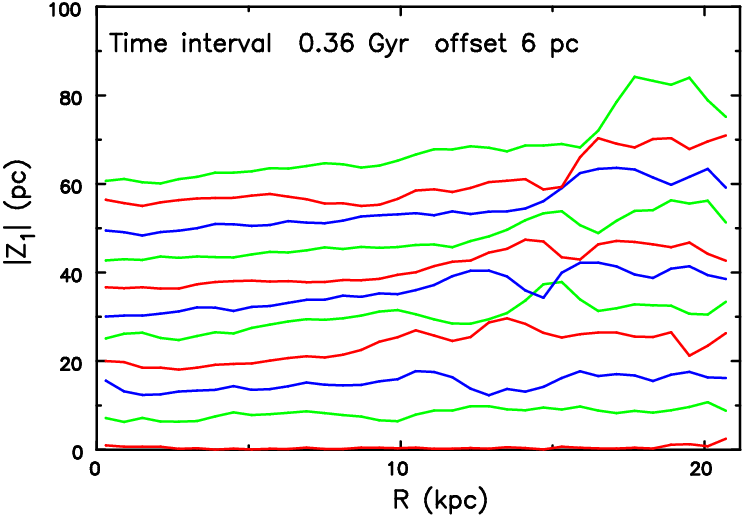}
\caption{The radial variation of $|Z_1|$ at equal time intervals in
  our standard run.  Successive lines, which are drawn at intervals of
  $0.36\;$Gyr, are shifted upward by 6pc and are coloured to make them
  easier to trace, and values based on fewer than 1000 particles are
  skipped.  This integral sign warp reaches a peak amplitude of $\sim
  24\;$pc near the outer edge of the disc by the last moment shown
  measured at $t=3.6\;$Gyr, which is the eleventh line.}
\label{fig.warp5059}
\end{figure}

Our first result, presented in Figure~\ref{fig.warp5059}, reveals
qualitatively similar behaviour to that reported by CW17, although
there are significant quantitative differences.  The salient feature
is that a small warp develops spontaneously, as CW17 reported, even
though the disc is almost completely stable to in-plane disturbances
and no external perturbations are applied.  By the last moment shown,
at 3.2~Gyr, the warp near the outer disc edge reaches an amplitude of
$\ga 20\;$pc over the radial range $17 < R <20\;$kpc, whereas CW17
reported that the warps in their models reached an amplitude of $\sim
100\;$pc by 4~Gyr and 300-400~pc by 10~Gyr.  CW17 suggested that their
warps propagated outwards across the disc, and there is a hint from
Figure~\ref{fig.warp5059} that may also happen in our model.  Note
also that the $|Z_1|$ displacements in the inner part of the disc in
Figure~\ref{fig.warp5059} have a tendency to rise almost linearly,
suggesting that the inner disc could be tilting rigidly, as we confirm
later, while the outer disc seems to flap less coherently.

Our model has a different bulge and halo from either of the models
used by CW17; we employed four times as many particles for the disk
and bulge and twice the number of halo particles, and we have used a
grid-based code whereas they used a tree code.  Also the $|Z_1|$
displacements reported in Figure~\ref{fig.warp5059} are relative to
the original disc plane of our model, whereas the warps reported by
CW17 were relative to the plane of the inner disc.

The warp in Figure~\ref{fig.warp5059} is of smaller amplitude than
that reported by CW17 for two reasons: first, they compute the
evolution of their models for approximately three times as long,
during which time the amplitude in the outer disc continued to
increase and, second, we employed more particles.  When we employed
one quarter the number of particles in each component, we found that
the warp amplitude reached $\sim 100\;$pc by 3.6~Gyr in reasonable
agreement with their value at 4~Gyr.  Furthermore, we recomputed the
same initial file of particles of our smaller $N$ model using a tree
code \citep[{\tt PKDGRAV}][]{Stadel01} with a softening length of
50~pc, finding a good quantitative agreement to $t=4\;$Gyr with the
results from the grid code, and also with CW17.

\begin{figure}
\includegraphics[width=.9\hsize,angle=0]{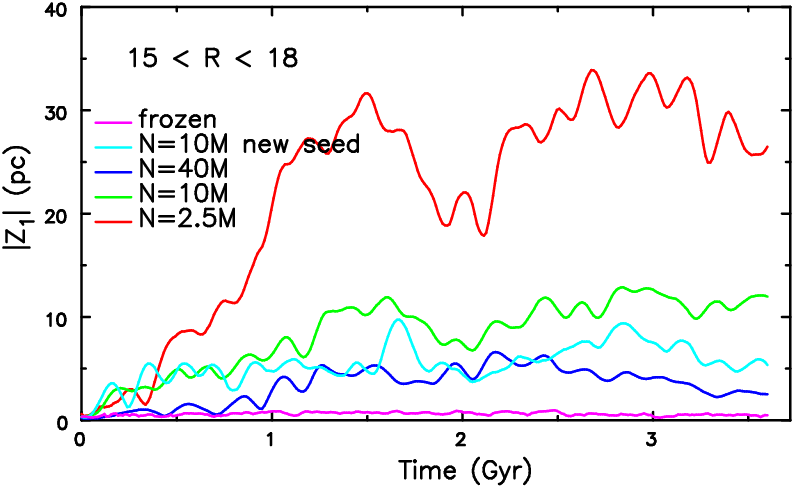}
\caption{The time evolution of the mass-weighted average of $|Z_1|$
  over the radial range given in kpc in five simulations.  The green
  line is from our standard case (Figure~\ref{fig.warp5059}), for
  which $N_{\rm halo}=10$M, and the numbers of particles employed in
  each mass component was reduced (red) and increased (blue) by
  factors of four.  The magenta line shows the effect of freezing the
  halo and bulge, while we simply changed the random seed in our
  standard run to obtain the result shown by the cyan line.}
\label{fig.live-froz}
\end{figure}

\begin{figure}
\includegraphics[width=.9\hsize,angle=0]{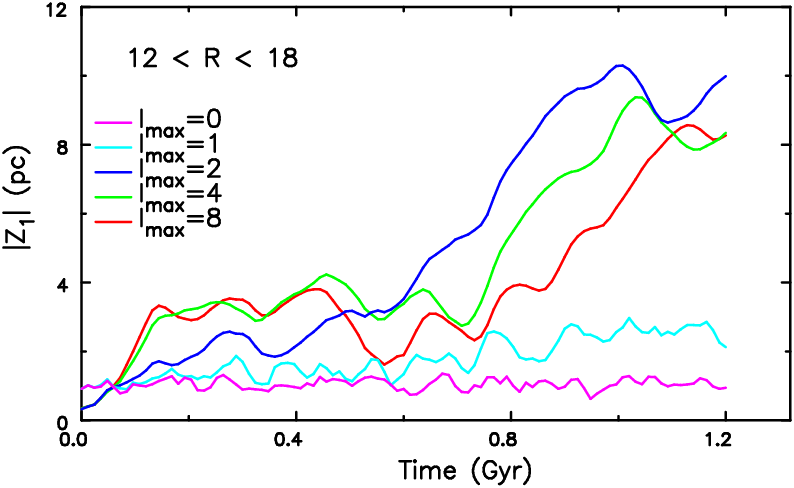}
\caption{As for Figure~\ref{fig.live-froz}, the early evolution of
  five simulations identical with our standard run, but in which we
  varied the order of expansion when determining the gravitational
  field of the halo and bulge particles.  The $Z_m$ values were
  averaged over the given radial range in kpc.  Again the green line
  is from our standard case (Figure~\ref{fig.warp5059}), for which
  $l_{\rm max}=4$.}
\label{fig.lmax}
\end{figure}

\subsection{Excitation mechanism}
\label{sec.hnoise}
We both reduced and increased the numbers of particles in each mass
component by factors of four, to obtain the red and blue curves shown
in Figure~\ref{fig.live-froz}, which reveal that the warp amplitude,
which fluctuates a great deal with time, generally seems to decrease
with increasing $N$.  Note that the warp amplitudes reported in this
figure are mass-weighted averages over the indicated radial range.
Freezing the halo prevents any warp from developing (magenta line),
and changing the initial random seed (cyan line) leads to quite
different evolution.  This evidence indicates that the warp is excited
by shot noise from the halo particles, which is reduced by increasing
$N$, eliminated by freezing the halo, and is subject to stochastic
variations as the noise spectrum is changed by a new random seed.

Figure~\ref{fig.lmax} presents the effect on the disc warp of changing
the number of spherical harmonics used to compute forces from the halo
and bulge particles; note these are shorter runs and the vertical
scale differs from that in Fig.~\ref{fig.live-froz}.  Again our
standard case, with $l_{\rm max}=4$, is shown by the green line, but
increasing the order of expansion to $l_{\rm max}=8$ (red line) made
only a small difference.  The warp behaviour was little different from
that in a frozen halo when we eliminated all aspherical terms (magenta
line), although in this case the halo particles were moving, and
therefore the monopole terms would have been subject to shot noise
fluctuations.  The dipole term (cyan line) excites a very mild warp,
but the quadrupole term ($l=2$) is clearly dominant (blue line), and
the warp grows more rapidly when $l_{\rm max}=2$ than it does when
higher multipoles are included.  It should also be noted that we tried
filtering out the $l=1$ and $l=3$ terms, while retaining $l=0$, 2 and
4, obtaining a result that was virtually indistinguishable from the
green curve for which the odd-$l$ terms were included.

The evidence in Figure~\ref{fig.lmax} is that the dominant component
that excites the warp is the quadrupole term in the expansion for the
gravitational field of the halo.  This seems physically reasonable,
since a quadrupole field that is misaligned with the plane of the disc
will attract the disc upward on one side and downward on the other,
exciting an $m=1$ warp in the disc.

The torque from shot noise acting on the disc is feeble and changes in
both amplitude and direction with time as the halo particles move on
their orbits.  Some density fluctuations excite the warp and,
presumably, others damp it, so the warp amplitude evolution is partly
a random walk caused by the evolving changes in the halo that apply a
changing torque on the disc.  Stochastic variations in the torque
acting on the disc must be partly responsible for the non-smooth time
variation of the bending amplitude (Fig.~\ref{fig.live-froz}).

\begin{figure}
\includegraphics[width=\hsize,angle=0]{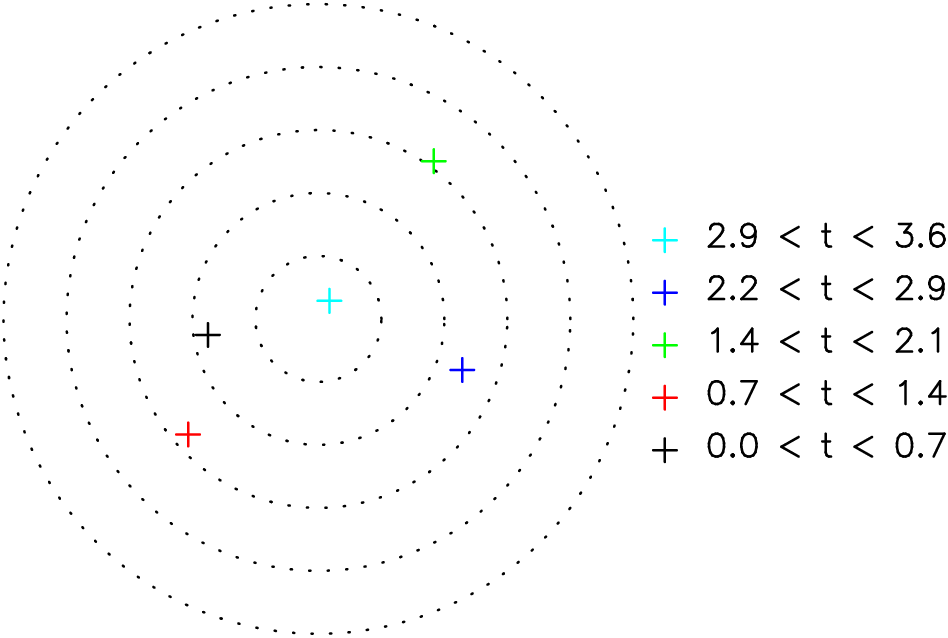}
\caption{The time evolution of the halo quadrupole orientation,
  $(\phi,\theta)$, estimated from the particles in the 6th energy bin
  of the halo that have a mean radius of $\sim15\;$kpc.  Each coloured
  point shows the average orientation during the period in Gyr
  indicated in the right margin.  Polar angles are relative to the
  stationary simulation frame; the radial coordinate is $\theta$ with
  the center being on the symmetry axis, $\theta=0$, and the outermost
  circle, $\theta=\pi/2$, is in the mid-plane, while the azimuth is
  $\phi$.  Only the upper hemisphere is shown, as the quadrupole has
  two-fold symmetry.}
\label{fig.Ylmdir}
\end{figure}
\begin{figure}

\includegraphics[width=.9\hsize,angle=0]{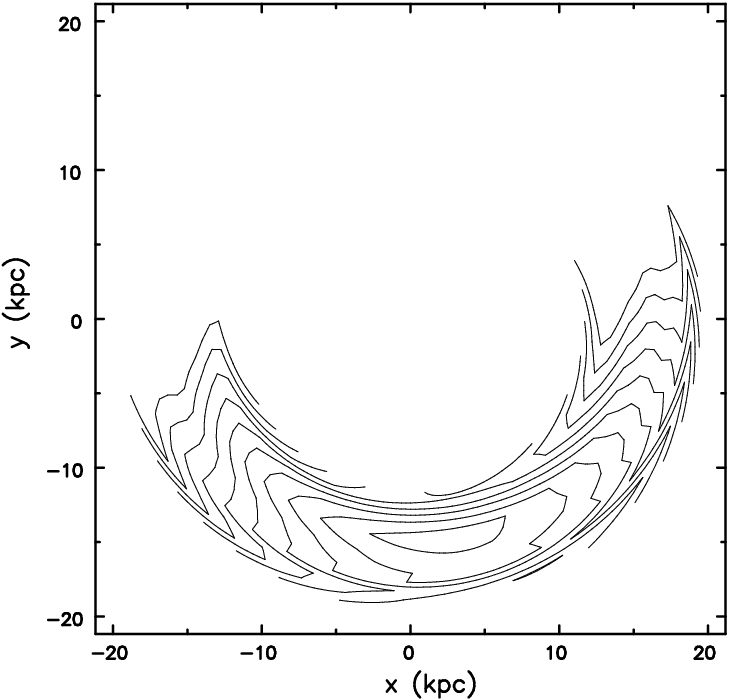}
\caption{The tilt of an annulus of the disc averaged over the interval
  from the start to 0.72~Gyr, in response to the torque from the black
  symbol in Figure~\ref{fig.Ylmdir}, which was computed from a similar
  radial range.  Only positive $Z_1$-displacements are shown.  Note
  that the disc tilt is approximately at right angles to the applied
  torque, as required by the vector cross product, since the disc
  rotates counter-clockwise in this projection.}
\label{fig.tip5059}
\end{figure}

To investigate this more fully, we divided the halo in our standard
run into 10 energy bins, each containing 1M particles, and computed a
spherical harmonic expansion, to $l_{\rm max} = 4$, of the density
distribution in each shell.  Tightly bound particles have the shortest
orbit periods and the direction of the quadrupole measured in the
innermost shells changes rapidly.  But we found some temporal
coherence in the orientation of the quadrupole in higher $E$ bins,
albeit with considerable jitter.  Figure~\ref{fig.Ylmdir} shows the
orientation of the quadrupole in the 6th energy bin, averaged over the
time intervals tabulated on the right.

The initial orientation in this energy bin, black symbol, lies near
$\phi \simeq 190^\circ$ with $\theta \simeq 32^\circ$, suggesting that
the halo attracts the disc upward near this value of $\phi$.  Because
the disc is rotating, the first tilt of the disc plane should be in
the direction given by the vector cross product of the disc angular
momentum and the applied torque, which is approximately true as shown
in Figure~\ref{fig.tip5059}.  The radial range shown in the disc was
chosen to match that of the halo particles used to compute the torque
orientation in Figure~\ref{fig.Ylmdir}, while the time interval
corresponds to that shown by the black symbol.

\begin{figure}
\includegraphics[width=.9\hsize,angle=0]{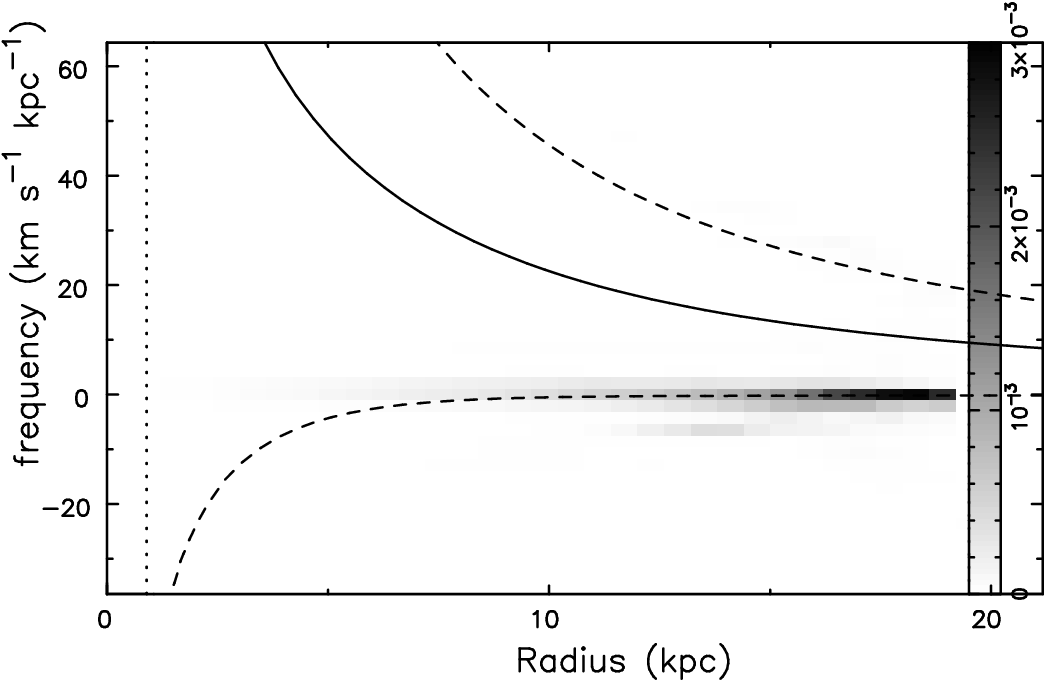}
\caption{A spectrum showing power in the $Z_1$ distortions of the disc
  as functions of frequency and radius in our standard simulation.
  The time interval extends over almost the entire evolution, and the
  radial range excludes the centre (inside the dotted line) and the
  outer edge of the disc (where the grey scale begins).  The solid
  curve marks the radial variation of the circular frequency
  $\Omega_c$, and the dashed curves $\Omega_c \pm \nu_{\rm ext}$, with
  $\nu_{\rm ext}$ being the vertical oscillation frequency caused by
  the attraction of the bulge and halo, \ie\ neglecting the self-force
  of the disc.}
\label{fig.spct5059}
\end{figure}

\subsection{Properties of the warp}
\label{sec.warp}
The initial forcing of the disc by the weak halo torques tilt the
plane of the disc as a rigid body in the inner parts, where the high
surface density and random motion make the inner disc stiff
\citep[\eg][]{DS99, SS06}.  These stiffening factors weaken with
radius, and the outer disc from $R \ga 4R_d$ \citep{SS06} is not
constrained to tip with the massive inner disc.  These statements are
supported by the evidence in Figure~\ref{fig.warp5059}.  As soon as
the outer disc becomes misaligned with the inner, the dominant torque
on it is no longer from the weak noise torques of the nearby halo,
but the coherent torque from the misaligned inner disc.  This strong
torque causes the outer disc to precess in the retrograde direction
\citep{SS06, Sell13}, while the massive inner disc barely moves in
response to the mild back reaction from the light outer disc.

All parts of the disc continue to be subject to torques from the halo
noise which must cause the inner disc tilt to change slowly, but the
dominant behaviour remains a barely moving inner disc that drives a
retrograde precession of the outer disc.  This behavior is confirmed
by the power spectrum of the $Z_1$ distortions in the disc over almost
the entire evolution of our standard model in
Figure~\ref{fig.spct5059}.  It reveals an almost zero frequency
disturbance over most of the disc, together with a retrograde
disturbance near the outer edge.  The frequencies of self-gravitating
warps are expected to be farther from corotation than $\Omega_c \pm
\nu_{\rm ext}$, and the slow wave to have a retrograde pattern speed
\citep{Sell13}.  Here $\nu_{\rm ext}$ is the vertical frequency
determined by the attraction of the bulge and halo, neglecting that of
the disc, as defined in \S\ref{sec.bends}.\footnote{The locations of
  warp resonances in a thickened disc depend on $\nu_{\rm ext}$ only,
  and not the total vertical frequency $\nu = (\nu_{\rm ext}^2 +
  \nu_{\rm int}^2)^{1/2}$ as was incorrectly stated by
  \citet[][\S8.1]{Sell13}.}

\begin{figure}
\includegraphics[width=.9\hsize,angle=0]{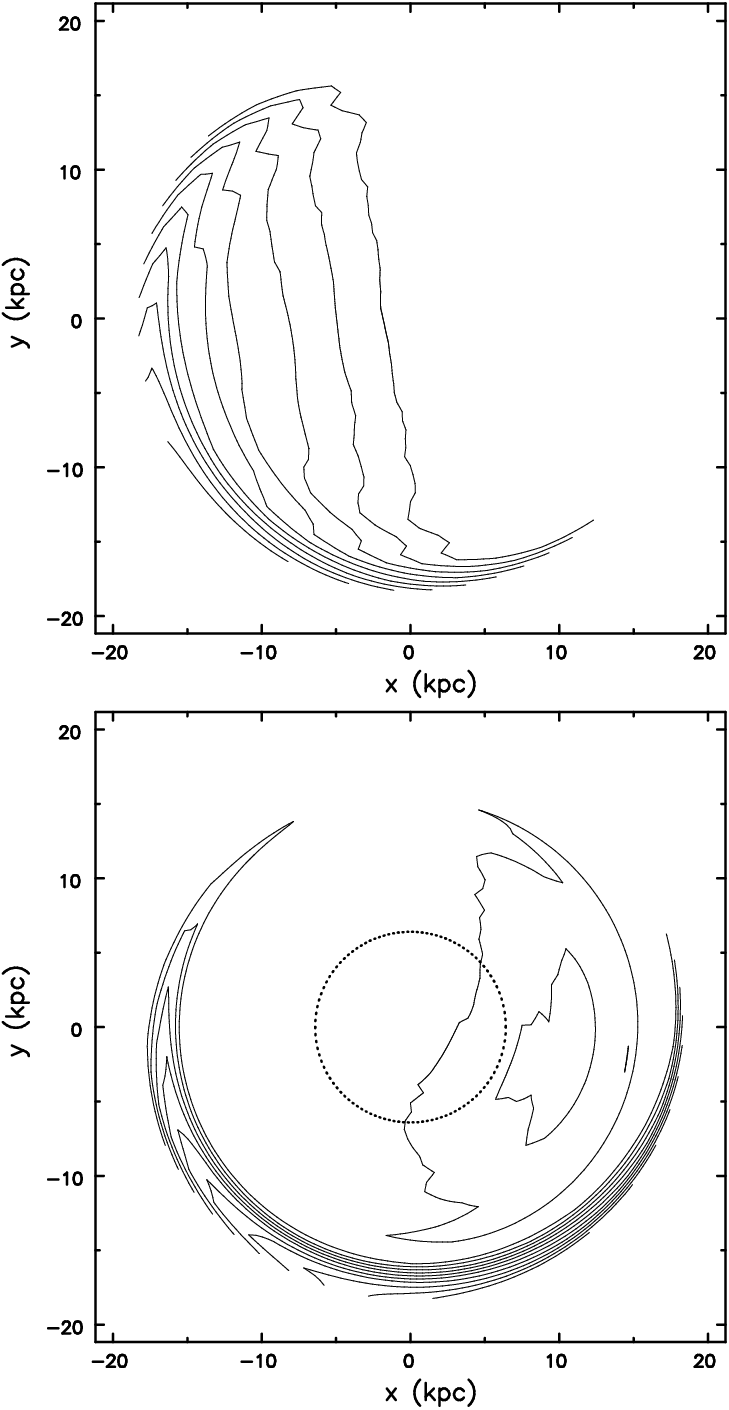}
\caption{The two best fitting disturbances to the $Z_1$-displacement
  data from our standard model.  The fit picks out time averages of
  the $Z_1$-displacement amplitude of possibly steadily rotating waves
  over the interval from 1.2~Gyr to 3.6~Gyr, and only positive
  $Z_1$-displacements are shown.  The dotted circle marks the radius
  at which $\Omega_p=\Omega_c-\nu_{\rm ext}$ for the retrograde
  rotating wave.}
\label{fig.modes5059}
\end{figure}

We have used the fitting code of \citet{SA86} to reveal the nature of
these two disturbances and display the results in
Figure~\ref{fig.modes5059}.  The top panel shows that the disturbance
of near zero frequency is indeed a simple tilt of the disc, as hinted
in Fig.~\ref{fig.warp5059}, whereas the retrograde disturbance has a
leading spiral shape, as also expected for slow warp waves
(\S\ref{sec.bends}).  We have not seen any evidence for exponential
growth, and so regard the warp as a driven response, not an
instability.  We find that the orientation of the inner disc tilt does
change very slowly over time, and that shown in top panel of
Figure~\ref{fig.modes5059}, which is drawn at the last fitted moment,
differs from that shown at the much earlier time in
Figure~\ref{fig.tip5059}.

We made similar fits to the data from the models of different $N$,
fitting two neutrally stable modes over $1.2 < t < 3.6\;$Gyr.  In each
case the fitted disturbances look very similar to those in
Figure~\ref{fig.modes5059}, and the frequencies are almost the same,
although the amplitudes differ, as expected.  

We do not suggest the leading spiral is a real mode, but rather a time
average of a slowly-evolving response driven by the misaligned inner
disk.  The group velocity of a leading bending wave is radially
outward (eq.~\ref{eq.vgroup}); not only does it slow as the disk
surface density decreases, but action conservation requires its
amplitude also to rise.

\begin{figure}
\includegraphics[width=.9\hsize,angle=0]{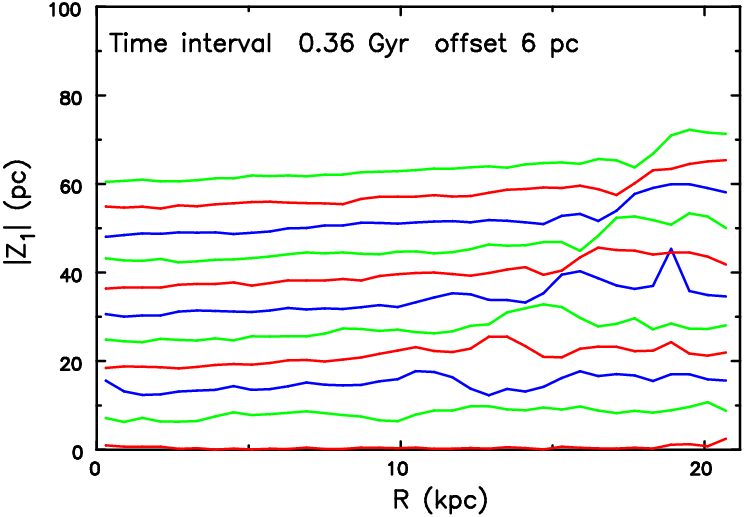}
\includegraphics[width=.9\hsize,angle=0]{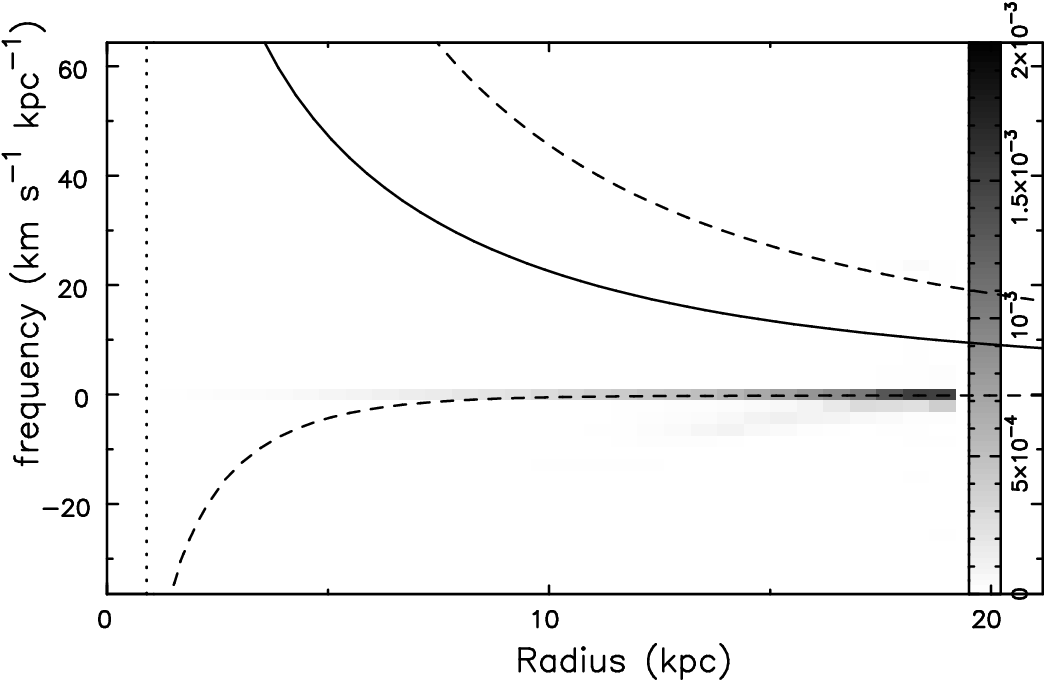}
\caption{Above: as for Fig~\ref{fig.warp5059}, which was drawn for our
  standard run, but for the new case in which the halo was frozen
  after 0.72~Gyr.  Although the warp is smaller, both the uniform tilt
  of the inner disc and the outward propagation of the bending wave
  are much clearer.  Below: the power spectrum of $m=1$ distortions in
  the new run with the frozen halo over the same time interval as
  shown in for the live halo in Fig.~\ref{fig.spct5059}.}
\label{fig.late-froz}
\end{figure}

\begin{figure}
\includegraphics[width=.9\hsize,angle=0]{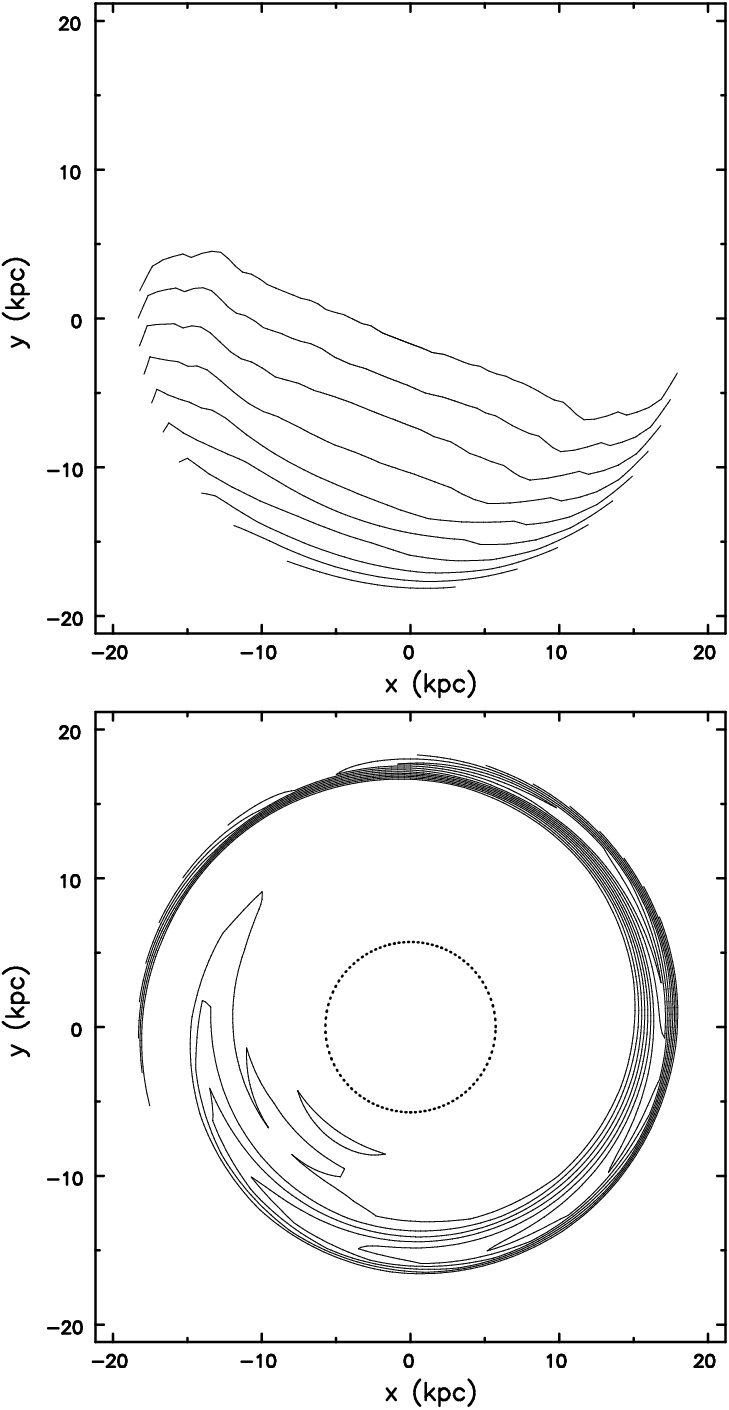}
\caption{The two best fitting disturbances to the $Z_1$-displacement
  data from the late frozen model, to be compared with the similar fit
  over the same time interval in our standard run with the live halo
  in Fig.~\ref{fig.modes5059}.}
\label{fig.modes5105}
\end{figure}

\subsection{A further test}
In order to confirm that the later evolution of the warp is largely
driven by disc dynamics, and is less affected by the stochastic torque
from the halo, we recomputed the later evolution of our standard run
(Fig.~\ref{fig.warp5059}) in a frozen halo.  After 0.72~Gyr, when the
inner disc had been tilted by the halo torque, we replaced the
live halo and bulge by rigid, smooth spherical mass distributions
having the same density profiles so that the disc equilibrium was
unaffected.

The time evolution of the warp in the new late-frozen halo case is
shown in the upper panel of Figure~\ref{fig.late-froz}, which is to be
compared with our standard result in Fig.~\ref{fig.warp5059}.  The
first three lines, at $t=0$, 0.36 and 0.72 Gyr are identical, of
course, but the later amplitude of the warp is lower in the frozen
halo case and fluctuates less -- presumably because the disk is no
longer being jerked around by on-going halo noise.  Note that both the
uniform tilt of the inner disc and the outward propagation of the
bending wave are much clearer in Figure~\ref{fig.late-froz} than in
Fig.~\ref{fig.warp5059}, and the radial extent of the uniformly tilted
disc grows as the bending wave propagates outward.  The power spectrum
from the late-frozen run is shown in the lower panel, which again
reveals a near stationary disturbance and a faint retrograde wave in
the outer disc.  These were the principal features in the comparison
case with the live halo (Fig.~\ref{fig.spct5059}), but the retrograde
wave is less pronounced in the late-frozen case.

The two fitted waves in Figure~\ref{fig.modes5105} also closely
resemble those in Figure~\ref{fig.modes5059}, with the leading spiral
shape being somewhat clearer in the lower panel of
Figure~\ref{fig.modes5105}.  This difference, and the different
orientations of the two disturbances, presumably arise because the
inner disc is not subject to on-going halo noise in the late-frozen
case.

This behaviour when the bulge and halo are rigid, smooth, spheres
confirms that the outwardly propagating wave in the outer disc can
be driven only by the torque from the inner disc.

\section{Conclusions}
Our principal conclusion is that a misalignment between the inner and
outer disc drives an outwardly-propagating bending wave that grows in
amplitude through action conservation and persists for several Gyr.
As the wave propagates outwards, it brings the previously misaligned
outer disc into alignment with the inner disc.  This behaviour, which
was also reported by \citet{SS06}, does not depend on the cause of the
misalignment or on its amplitude -- their simulation revealed a much
larger warp that behaved in exactly the same way as the mild features
we report here.

The disc is a flywheel with a huge store of angular momentum, and the
inner part is quite stiff.  The inner disc therefore responds to any
externaly applied torque by tipping slightly, while the outer part may
tip more.  As soon as a misalignment between the inner and outer disc
is established, it launches an outwardly travelling, retrograde,
leading spiral bending wave.  We showed here that this behaviour
persisted in a case in which we froze the halo after some evolution,
demonstrating that the dynamical mechanism is indeed dominated by the
misalignment between the inner and outer disc.

We have confirmed the result from CW17 that mild warps are excited
stochastically by halo noise.  The supporting evidence is that we find
warp amplitudes decrease in experiments having larger numbers of halo
particles, which have milder density fluctuations, and changing the
random seed leads to a completely different warp history.  We have
developed this picture by showing that the warp is excited most
effectively by the quadrupole component of the halo noise.  Since the
quadrupole can have any orientation, which will generally be inclined
to disc plane, it naturally excites an $m=1$ tip to the disc.

All the warps reported in this paper are excited by the very mild shot
noise in our carefully-constructed equilibrium halo, and such small
warps would be barely observable.  Halo substructure in real galaxies,
on the other hand, as well as any residual sloshing motions from
galaxy assembly, will exert larger perturbing forces that would also
be more temporally coherent as the sub-halos move on their orbits.

\section*{Acknowledgements}
JAS acknowledges the continuing hospitality of Steward Observatory.
VPD is supported by STFC Consolidated grant \#~ST/R000786/1.

\section*{Data availability}
The data from the simulations reported here can be made available
on request.  The simulation code can be downloaded from
{\tt http://www.physics.rutgers.edu/galaxy}


\bsp	
\label{lastpage}
\end{document}